\documentclass[a4paper,11pt]{article}
\usepackage{moriond,epsfig}
\usepackage{graphicx}
\usepackage{bm}
\usepackage{multirow}
\usepackage{amsmath}
\usepackage{amssymb}
\usepackage{subfigure}

\def\jpsi{J/\psi}

\def\be{\begin{equation}}
\def\ee{\end{equation}}
\def\bea{\begin{eqnarray}}
\def\eea{\end{eqnarray}}

\begin{document}                                                               
\vspace*{3.8cm}
\title{Search for Excess Dimuon Production in the Radial Region $\bm{1.6 < r}$ $\lesssim$ $\bm{10}$~cm at the D0 Experiment}

\author{ M.R.J. Williams \\ Lancaster University}
\address{Fermilab, Batavia, Il 60614, U.S.A.}

\maketitle\abstracts{
We report on a study of dimuon events produced in $p\bar{p}$ collisions at $\sqrt{s} = 1.96$~TeV, using $0.9$~fb$^{-1}$ of data recorded by the D0 experiment during 2008.
Using information from the inner-layer silicon tracking detector, we observe $712 \pm 462 \pm 942$ events in which one or both muons are produced in the range $1.6 < r \lesssim 10$~cm, which is expressed as a fraction $(0.40 \pm 0.26 \pm 0.53)\%$ of the total dimuon sample. We therefore see no significant excess of muons produced a few centimeters away from the interaction point.
}

\section{Introduction}

A recent study by the CDF collaboration~\cite{ghosts} claims to observe a large sample of dimuon events in which one or both muons appear to be produced at large radial distances ($> 1.5$~cm) from the primary interaction point, with observations not consistent with current heavy quark production models. We describe a corresponding search at D0 for dimuon events in which one or both muons are produced at radial distances exceeding $1.6$~cm, relative to the primary $p\bar{p}$ interaction.

\section{\label{sec:d0}The D0 Detector}

D0 is a general purpose colliding beam detector described in detail elsewhere~\cite{d0det}. The most important detector components for this analysis are the tracking and muon systems in the central region corresponding to pseudorapidity $|\eta| < 1.0$.

The central tracking detector at D0 is contained within a $2$~T solenoidal field, and comprises the silicon microstrip tracker (SMT) and the central fiber tracker (CFT). The silicon tracker consists of six barrel/disk modules, two outer disks, and an inner layer (L0). The L0 detector~\cite{L0} surrounds the beryllium beampipe, and is formed from single-sided sensors staggered at radii of 1.60 and 1.76~cm to provide 98\% $\phi$ coverage, in the range $|z| < 38$~cm.

The muon system in the $|\eta| < 1$ region consists of a combination of proportional drift tubes (PDTs) and scintillation counters~\cite{d0muon}. These are arranged on both sides of an iron toroid of thickness $1.09$~m, which provides a magnetic field of $1.8$~T to aid in muon identification and reconstruction. The total thickness traversed by muons exiting the iron is 12.8--14.5 $\lambda$ giving a minimum momentum of around $3$~GeV/$c$ for that muon topology.

\section{\label{sec:select}Event Selection}

This analysis uses data collected by the D0 experiment between August and December 2008, corresponding to a total integrated luminosity of around 0.9~fb$^{-1}$. No particular trigger requirements are enforced (i.e. events are accepted from all triggers, inclusively). The event selection scheme is designed to approximately match the corresponding requirements used by CDF in their analysis~\cite{ghosts}. 

The dimuon sample is produced by selecting the two highest-$p_T$ muons in each event, provided that they satisfy the following requirements. Both muons must fulfill standard D0 quality criteria, be associated with muon-system hits on both sides of the toroid magnet, and satisfy $p_T > 3$~GeV/$c$ and $|\eta| < 1.0$. Both muons must be matched to central tracks, which originate within $1.5$~cm of each other along the beam axis. The combined invariant mass of the dimuon pair, $M(\mu\mu)$, must lie in the range $5 < M(\mu\mu) < 80$~GeV/$c^2$. To remove contamination from cosmic ray muons, opposite-sign dimuon candidates are excluded if their azimuthal separation $\Delta\phi$ exceeds $3.135$~rad. Both muons must also be detected within $\pm 10$~ns of the expected arrival time (for beam-produced muons) by muon scintillators on either side of the toroid. Finally, muons are excluded if they do not pass through the active geometrical limits of the L0 detector.

Two subsets of the dimuon sample are defined using information from the SMT. Muons are tagged as ``loose'' if they have three or more hits in the silicon tracker, and tagged as ``tight'' if they are also associated with a L0 hit. A loose (tight) \emph{event} is one in which both muons are loose (tight). Events which are loose but not tight therefore contain at least one muon without a L0 hit, either due to hit inefficiencies, or production beyond L0. The loose selection has a radial range up to $10$~cm, determined by examining the hit coordinates of muons accepted in this sample.

The number of events containing one or both muons produced beyond L0, $N(\text{excess})$, can be expressed in terms of the number of loose and tight events, and the dimuon selection efficiency of the tight sample relative to the loose, $\varepsilon_{T/L}$:
\begin{eqnarray}
N(\text{excess}) = N(\text{loose}) - \frac{N(\text{tight})}{\varepsilon_{T/L}}~.
\label{eq:count} 
\end{eqnarray}


\section{\label{sec:eff}Tight/Loose Muon Detection Efficiencies}

To measure the relative efficiency of the tight and loose selection requirements, a test sample of $\jpsi \to \mu^+ \mu^-$ candidates is selected. The same 0.9~fb$^{-1}$ data sample described above is used to reconstruct $\jpsi$ candidates from opposite-charge muons originating at a common vertex, and with invariant mass $2.95 < M(\mu\mu) < 3.2$~GeV/$c^2$. All other requirements applied to the signal sample are also applied to the $\jpsi$ muons. The test sample is limited to muons produced within L0 by enforcing the transverse decay length criterion $|L_{xy}(\jpsi)| < 1.6$~cm, which is the case for $>99.89$\% of the selected candidates.

The resulting test sample contains $119\,276$ $\jpsi \to \mu^+ \mu^-$ candidates, of which $(91.5 \pm 0.6)\%$ are signal $\jpsi$ events, as determined by fitting the dimuon invariant mass distribution with an appropriate model. From this sample of $238\,552$ muons, $228\,569$ satisfy the loose SMT criteria, and $210\,026$ also satisfy the tight (L0) requirement. The mean single-muon efficiency, $\varepsilon_{T/L}(\mu)$, is therefore calculated to be $0.9189 \pm 0.0006$; the equivalent dimuon efficiency is the square of this value, $\varepsilon_{T/L} = 0.8443 \pm 0.0008$. Here the uncertainties are purely statistical, and are determined using the standard binomial expression. 

\begin{figure}[t]
        \centering
	\subfigure[$p_T$-dependent tight/loose selection efficiency, showing the result of a fit to a linear function (solid line), and the mean value (dashed line).]
		{\label{fig:jpsi_pt_eff}\includegraphics[width=7.0cm]
                          {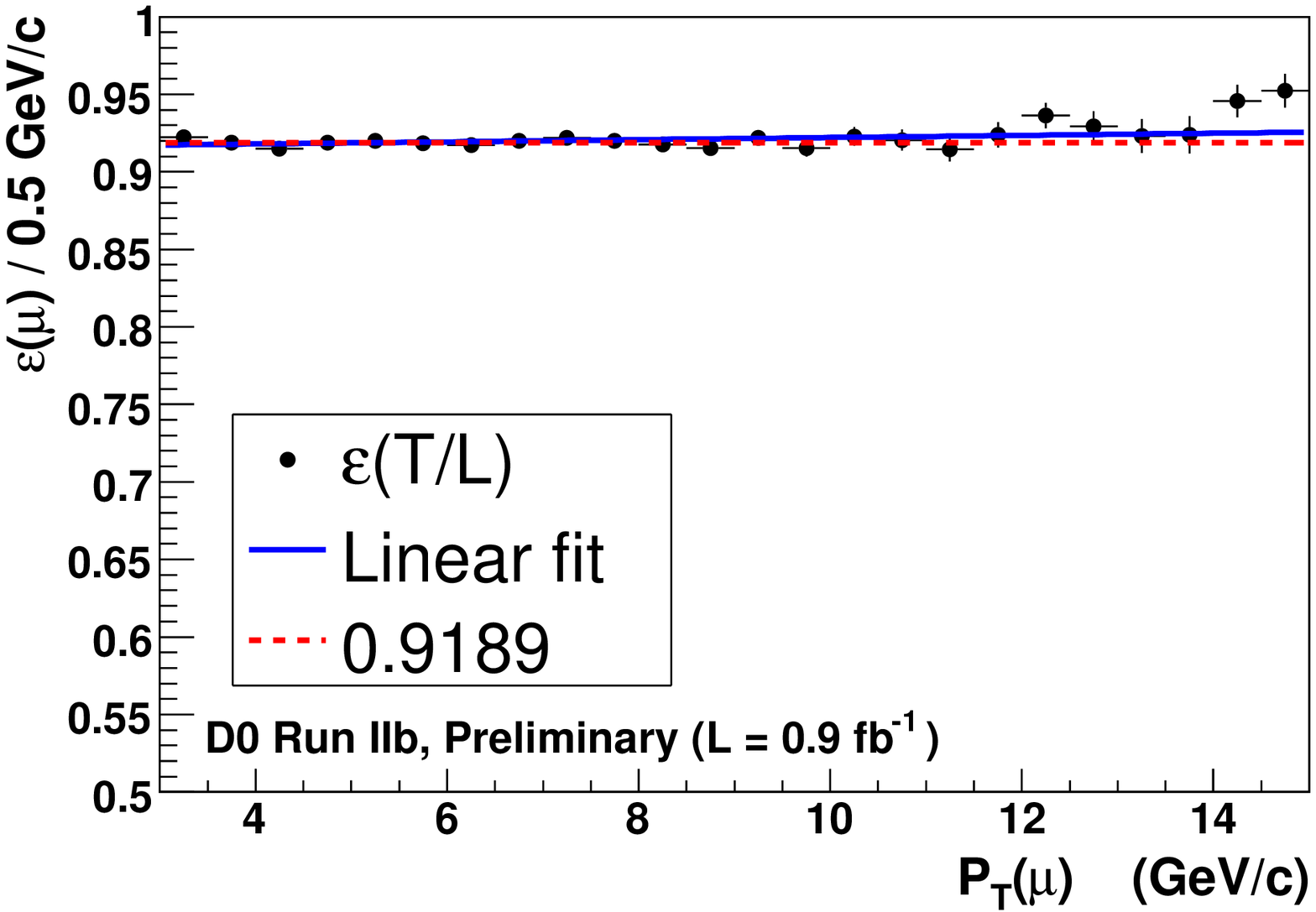}}
	\subfigure[Two-dimensional $(z,\phi)$-dependent tight/loose selection efficiency. The low bins in $z$ are associated with the $\sim$$1$~mm gaps between the sensors.]
		{\label{fig:jpsi_phi_z_eff}\includegraphics[width=7.0cm]
                          {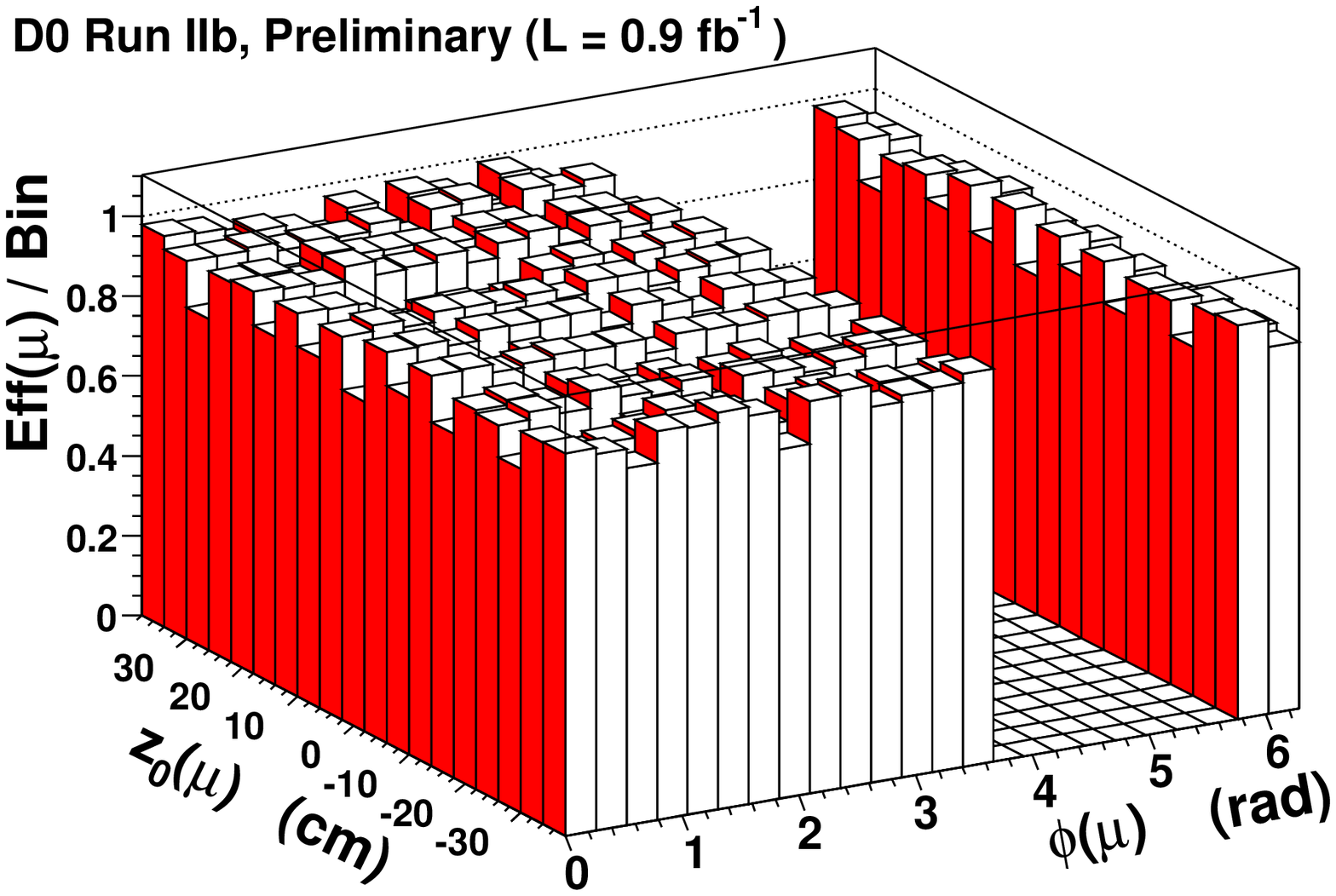}}
        \caption[]{The relative tight/loose selection efficiency for muons from $\jpsi$ decays, as a function of their transverse momentum $p_T$ and two-dimensional geometry $(z,\phi)$.}
\end{figure}

The efficiency $\varepsilon_{T/L}(\mu)$ can depend on the kinematic and geometrical properties of the muon. Figure~\ref{fig:jpsi_pt_eff} shows the relative tight/loose selection efficiency as a function of transverse momentum, determined by calculating $N(\text{tight})/N(\text{loose})$ separately in each $p_T$ bin. The distribution suggests that higher-$p_T$ muons may have higher L0 efficiencies, at the $\sim$$1\%$ level, which is quantified by fitting the resulting distribution to a first degree polynomial, yielding the parameterization $\varepsilon(p_T) = a + b \cdot (p_T - 9)$, where $a = 0.9214 \pm 0.0011$, $b = 0.0007 \pm 0.0003$, and the transverse momentum is in units of GeV/$c$. The fit converges with a minimum $\chi^2$ of 31, for 22 degrees of freedom.

A similar calculation is carried out for the muon pseudorapidity $\eta$. In this case, small ($\pm 3\%$) and uncorrelated variations in efficiency are observed in different bins, and are taken into account on a bin-by-bin basis as described later. Finally, the efficiency is determined as a function of the muon $(z,\phi)$ coordinate, as it passes through the L0 detector, as shown in Fig.~\ref{fig:jpsi_phi_z_eff}. This accounts for the effects of varying sensor performance.

The relative efficiency of loose and tight requirements for dimuon events is then taken as the product of the efficiencies for the individual muons, where the individual muon efficiency is considered as the product of kinematic and geometrical contributions:
\begin{eqnarray}
\varepsilon(\mu_i)       & = & \varepsilon(z_{\mu_i},\phi_{\mu_i}) \cdot \mathcal{F}(\eta_{\mu_i}) \cdot \mathcal{F}(p_T^{\mu_i})~. \label{eq:efficiency2}
\end{eqnarray}
The geometrical efficiency $\varepsilon(z_{\mu_i},\phi_{\mu_i})$ is taken directly from the histogram shown in Fig.~\ref{fig:jpsi_phi_z_eff}. The normalized pseudorapidity dependence $\mathcal{F}(\eta_{\mu_i})$ is taken directly from the appropriate bin of the $\varepsilon(\mu_i)$ vs. $\eta$ histogram.
The normalized transverse momentum dependence is parameterized as:
\begin{eqnarray}
\mathcal{F}(p_T^{\mu_i}) & = & \frac{ a + b \cdot (p_T^{\mu_i} - 9)}{a}~,    \label{eq:efficiency21}
\end{eqnarray}
with $a$ and $b$ fixed at their central values, and transverse momentum in units of GeV/$c$. For muons with transverse momenta larger than $15$~GeV/$c$, the $p_T$ correction is fixed to unity.


\section{\label{sec:results}Results}

In total, the signal sample contains $204\,177$ dimuon events, of which $177\,535$ satisfy the loose silicon hit requirements. Of these, $149\,161$ also pass the tight SMT requirement that both muons have a hit in L0. The determination of $N$(excess) proceeds on an event-by-event basis:
\begin{eqnarray}
N(\text{excess})  &=&  N^{\text{obs}}(\text{loose}) -  \sum_{i=1}^{N(\text{tight})} \frac{1}{ \varepsilon^i_{T/L}} 
\label{eq:extrap1}
\end{eqnarray}
Here $N^{\text{obs}}$(loose) is the observed number of events in the loose sample. The summation is over all tight events, and the efficiency for an event is the product of the two muon efficiencies given by Eq.~(\ref{eq:efficiency2}). Using this method, the total number, and relative fraction, of events with one or both muons produced outside L0 is found to be:
\begin{eqnarray}
N(\text{excess}) & = &  712            \pm  462~ \text{(stat.)} \pm 942~ \text{(syst.)}, \label{eq:results1} \\
N(\text{excess}) / N^{\text{obs}}(\text{loose}) & = & ( 0.40 \pm  0.26 \pm 0.53) ~\% .  \label{eq:results1a}
\end{eqnarray}
This value is significantly smaller than the corresponding fraction of 12\% reported by CDF.

The uncertainty on the efficiency is determined using ensemble tests, in which the full event-by-event calculation of Eq.~(\ref{eq:extrap1}) is repeated 1000 times, with the constituent efficiency factors in Eq.~(\ref{eq:efficiency2}) allowed to vary pseudo-randomly with an appropriate mean and standard deviation. The resulting uncertainty $\pm 0.0024$ is then translated into an uncertainty on $N(\text{excess})$.


\section{\label{sec:syst}Systematic Uncertainties and Consistency Checks}

A systematic uncertainty is assigned to the efficiency calculation by repeating it with different binning schemes for the $\phi$, $z$, and $\eta$ histograms, and also with the $p_T$ factor removed from the parameterization. These sources combine to give a total systematic uncertainty on the fractional excess of $\pm 0.53 \%$. 

The study is repeated using only those events which are selected by a dedicated dimuon trigger in which the muons are not required to be matched to central tracks. The results are consistent with the inclusive trigger selection, with $N$(excess) comprising a fraction 0.32\% of the loose sample ($396$ / $123,572$).

To check for possible enhancements of $N$(excess) in different kinematic regions, it is determined separately in bins of $M(\mu\mu)$, $\phi(\mu)$, $\eta(\mu)$ and $p_T(\mu)$, and no significant departure from zero is observed in any particular region.

\section{\label{sec:conc}Conclusions}

Using $0.9$~fb$^{-1}$ of integrated luminosity, a sample of dimuon events is examined with similar event characteristics as those selected in the CDF multi-muon analysis~\cite{ghosts}. We measure the fraction of events in which one or both muons are produced in the range $1.6 < r \lesssim 10$~cm to be $(0.40 \pm 0.26 \pm 0.53)\%$, significantly smaller than the $\sim$$12\%$ observed by CDF. These events are expected to have contributions from $K$ and $\pi$ decays in flight, as well as residual cosmic ray contamination, and further studies are in progress to quantify the composition. Additional examination of the dimuon properties indicate no significant excess in any particular kinematic regions in $\eta$, $\phi$, $p_T$ or $M(\mu\mu)$.


\end{document}